# A MODEL FOR MINING MULTILEVEL FUZZY ASSOCIATION RULE IN DATABASE

Pratima Gautam
Department of computer Applications
MANIT, Bhopal

Neelu Khare
Department of computer Applications
MANIT, Bhopal (M.P.)

K. R. Pardasani
Department of Mathematics
MANIT, Bhopal (M.P.)

**Abstract:** The problem of developing models and algorithms for multilevel association mining pose for new challenges for mathematics and computer science. These problems become more challenging, when some form of uncertainty like fuzziness is present in data or relationships in data. This paper proposes a multilevel fuzzy association rule mining models for extracting knowledge implicit in transactions database with different support at each level. The proposed algorithm adopts a top-down progressively deepening approach to derive large itemsets. This approach incorporates fuzzy boundaries instead of sharp boundary intervals. An example is also given to demonstrate that the proposed mining algorithm can derive the multiple-level association rules under different supports in a simple and effective manner.

## 1 INTRODUCTION

Data mining, referred to as knowledge discovery from databases, a process of nontrivial extraction of implicit, previously unknown and potentially useful information from databases, has wide application in information management, query processing, decision making, process control, etc [1], [10]. An Association rule mining is an important process in data mining, which determines the correlation between items belonging to a transaction database [11], [13]. In general, every association rule must satisfy two user specified constraints: support and confidence. The support of a rule $X \Rightarrow Y$ is defined as the fraction of transactions that contain $X \cup Y$, where X and Y are disjoint sets of items from the given database [10], [15]. The confidence is defined as the ratio support $(X \cup Y)/support(X)$. Here the aim is to find all rules that satisfy user specified minimum support and confidence values. A good number of models and algorithms are reported in the literature [13] for single level association rule mining, which has been developed using techniques from mathematics, statistics and computer science. Some applications even need us to mine association rules in multiple levels of abstraction [2]. For example, a user may not only be concerned with the associations between "computer" and "printer", but also wants to know the association between desktop PC price and laser printer price. Some approaches for constructing a concept hierarchy and then trying to discover knowledge in the multi-level abstractions to solve this problem are reported in the literature [7] like Apriori algorithm [13], [10] etc. Under the same minimum support and minimum confidence thresholds. This method is simple, but may lead to some undesirable results. Infact different levels should have different support to extract appropriate patterns. Higher support usually exists at higher levels and if one wants to find interesting rules at lower levels, he/she must define lower minimum support values [4], [6].
Another trend to deal with the problem is based on fuzzy theory [11], [12] which provide an excellent means to model the "fuzzy" boundaries of linguistic terms by introducing gradual membership [3].Instead of using sharp boundary intervals, fuzzy sets are used in discovering fuzzy association rules, which are easily understandable to humans [2], [14],[15]. Here efficient model based on fuzzy sets and Han's mining approach for multiple-level items is proposed. It adopts a top-down progressively deepening method to finding large itemsets. It integrates fuzzy-set concepts, data-mining technologies and multiple-level taxonomy to find fuzzy association rules in given transaction data sets. The mined rules are more natural and understandable for human beings.

## 2. Apriori Algorithm and Apriori Property:

Apriori is an influential algorithm in market basket analysis for mining frequent itemsets for Boolean association rules. The name of Apriori is based on the fact that the algorithm uses prior knowledge of frequent itemset properties. Apriori employs an iterative approach known as a level-wise search, where k-itemsets are used to explore (k+1)-itemsets [9], [13]. First, the set of frequent 1-itemsets is found, denoted by L1. L1 is used to find L2, the set of frequent 2-itemsets, which is used to find L3, and so on, until no more frequent k-itemsets can be found.
Property: All non empty subsets of frequent item sets must be frequent.

### 2.1 Multilevel Association Rule:

In multilevel association rules mining, different minimum support setting should be used at different concept levels. We discover frequent patterns and strong association rules at the top-most concept level. Assume the minimum support at this level is 5 percent and the minimum confidence is 50 percent. One may find a set of single frequent items (each called a frequent 1-itemset), a set of pair-wised frequent items (each called a frequent 2-itemset) and a set of strong association rules [7], [8]. At the second level, if the minimum support is 2 percent and the minimum confidence is 40 percent, one may find frequent 1-itemsets and frequent 2-itemsets and strong association rules. The process repeats at even lower concept levels until no frequent patterns can be found [3]. During multi-level association rule mining, the taxonomy information for each (grouped) item in figure.1 is encoded as a sequence of digits in the transaction table T [1] represents, For example, the item dairyland chocolate milk' is encoded as `111' in which the first digit, `1', represents `milk' at level-1, the second, `1', for the flavor 'chocolate' at level 2, and the third, `1', for the brand `Dairyland' at level-3. Similar to [6], repeated items (i.e., items with the same encoding) at any level will be treated as one item in one transaction.

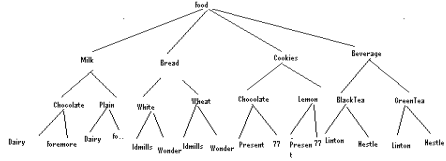

"(Fig. 1)": The taxonomy for the relevant data items

## 3 The Proposed Model:

In the proposed Model, items may have different minimum supports and taxonomic relationships to discover the large itemsets. Each level is given a predefined support threshold (minimum support). The minimum support for an itemset is set as the minimum supports of the items contained in the itemset, while the minimum support for an item at a higher taxonomic concept is set as the minimum supports of the items belonging to it. Under the constraint, the characteristic of downward-closure is kept, such that the original apriori algorithm can be easily extended to find the fuzzy large itemsets [2]. In this paper, we use Han and Fu's encoding scheme, as described in [7] to represent nodes in predefined taxonomies for mining multilevel rules. Nodes are encoded with respect to their positions in the hierarchy using sequences of numbers and the symbol "*". It then filters out unpromising itemsets in two phases. In the first phase, an item group is removed if its occurring count is less than the support threshold. In the second phase, the count of a fuzzy region is checked to determine whether it is large [5], [9]. The proposed algorithm then finds all the large itemsets for the given transactions by comparing the fuzzy count of each candidate itemset with its support threshold. Furthermore, some pruning strategies are used to reduce the number of candidate itemsets generated. The proposed algorithm is given in the following steps:

Step 1:
Encode taxonomy using a sequence of numbers and the symbol "*", with the $l$th number representing the branch number of a certain item at levels.

Step-2:
Determine $\gamma \in \{2, 3,....\}$ (Maximum item threshold). $\gamma$ is a threshold to determine maximum number of items in a transaction by which the transaction may or may not be considered in the process of generating rules mining. In this case, the process just considers all transactions with the number of items in the transactions less than or equal to $\gamma$. Formally, let D be a universal set of transactions. $M \subseteq D$ is considered as a subset of qualified transactions for generating rules mining that the number of items in its transactions is not greater than $\gamma$ as defined by:

$$M = \{T \mid card(T) \leq \gamma, T \in D\}, \quad (1)$$

Where card (T) is the number of items in transaction T.

Step-3:
Set k = 1, where k is used to store the level number being processed whereas k $\in$ {1, 2, 3} (as we consider up to 3-levels of hierarchies).

Step-4:
Set p=1, where p is an index variable to determine the number of combination of items in itemsets called p-itemsets.
p $\in$ {1,2,3} (as we consider up to 3-itemsets at each level of hierarchy).

Step-5
Determine minimum support for p-itemsets at level k, denoted by $\alpha_p^k \in (0, |M|]$ as a minimum threshold of a combination items appearing in the whole qualified transactions, where |M| is the number of qualified transactions. $\alpha_p^k$ May have same value for every p at level k.

Step-6
Group the items with the same first k digits in each transaction $T_j$, and add the occurrence of the items in the same groups in Ti. Denote the amount of the j-th group $I_j^p$ for $T_i$ as $v_{ij}^p$.

Step 7:
Construct every candidate p-itemset, $I^p$ as a fuzzy set on set of qualified transactions, M. A fuzzy membership function $\mu_{I^p}$ is a mapping: $M \rightarrow [0, 1]$ as defined by:

$$\mu_{I^p}(T) = v_{ij}^p \cdot \inf_{i \in I^p}\left\{\frac{\eta_T(i)}{Card(T)}\right\}, \forall T \in M \quad (2)$$

Where T be a qualified transaction in which T can be regarded also as a subset of items. $T \subseteq D$

A Boolean membership function, $\eta$, is a mapping

$$\eta_T : D \longrightarrow \{0,1\}$$

as defined by:

$$\eta_T(i) = \begin{cases} 1, i \in T \\ 0, otherwise \end{cases} \quad (3)$$

Such that if an item, i, is an element of T then $\eta_T(I) = 1$, otherwise $\eta_T(i)=0$.

Step - 8:
Calculate *support* for every (candidate) p-itemset using the following equations:

$$Support(I^p)^k = \sum_{T \in M} v_{ij}^p \cdot \mu_{I^p}^k \quad (4)$$

M is the set of qualified transactions; it can be proved that (4) satisfied the following property:

$$\sum_{i \in D} Support(i) = |M|$$

(5)

For $p=1$, $I^p$ can be considered as a single item. if $p>1$ then generate candidate set $C_2^k$ has to following steps for each newly from 2-itemsets.

Step-9:
$I^p$ will be stored in the set of frequent p-itemsets, $L_p^k$ if and only if $support(I_i^p) \geq \alpha_p^k$.

Step-10
Set $p=p+1$, for the same level k and if $p>3$, then go to Step-12.

Step-11
Looking for possible/candidate p-itemsets from $L_{p-1}$ by the following rules:
A p-itemset, $I^p$, will be considered as a candidate p-itemset if $I^p$ satisfies:
$\forall F \subset I^p | F |= K - 1 \Rightarrow F \in L_{p-1}$

For example, $I^p=\{411, 412, 422, 211, 221\}$ will be considered as a candidate 5-itemset, iff: $\{222, 411, 211\}$, $\{222, 212, 322\}$, $\{322, 212, 422, 311\}$ and $\{311, 412, 221, 311, 322\}$ in $L_3$. If there is no candidate p-itemset then go to Step-12. Otherwise, the process is going to Step-4

Step-12 Similar to Apriori Algorithm, confidence of an association rule mining, $A \Rightarrow B$, can be calculated by the following equation:

$$Conf(A \Rightarrow B) = P(B|A) = \frac{Support(A \cup B)}{Support(A)}$$

(6)

Where $A, B \in D$
It can be followed that (5) can also be expressed by

$$Conf(A \Rightarrow B) = \frac{\sum_{T \in M} \inf_{i \in A \cup B}(v_{ij}^p \cdot \mu_i(T))}{\sum_{T \in M} \inf_{i \in A}(v_{ij}^p \cdot \mu_i(T))}$$

(6)

Where A and B is any p-itemsets in Lp.   $[\mu_i(T) = \mu_{(i)}(T)]$
.Therefore, support of an itemset as given by (4) can be expressed as following:

$$Support(I^p)^k = \sum_{T \in M} \inf_{i \in I^p}(\mu_i(T))$$

(7)

Step -13
Set $k=k+1$ and go to step-6 (for repeating the whole processing for next level).

## 4. AN ILLUSTRATIVE EXAMPLE

An illustrative example is given to understand well the concept of the proposed model and algorithm and how the process of the generating fuzzy association rule mining is performed step by step. The process is started from a given transactional database as shown in Table 1[a].

TABLE 1[a]

| Trans_ID | List of items |
|---|---|
| T1 | 222, 411, 211 |
| T2 | 411, 412, 422, 211, 221 |
| T3 | 222, 212, 322, 412 |
| T4 | 311, 111, 212, 211, 411, 412 |
| T5 | 322, 212, 422, 311 |
| T6 | 322, 222, 312, 411 |
| T7 | 311, 412, 221, 311, 322 |
| T8 | 212, 411, 321, 221 |
| T9 | 311, 312, 421, 212, 211, 322 |
| T10 | 322, 122 |
| T11 | 222, 111 |
| T12 | 111 |

Table1 [b]
**Codes of item name**

| Item name (terminal node) | Code | Item name (internal node) | Code |
|---|---|---|---|
| Dairyland chocolate milk | 111 | Milk | 1** |
| Foremore plain milk | 122 | Bread | 2** |
| Old Mills white bread | 211 | Cookies | 3** |
| Wonder white bread | 212 | Beverage | 4** |
| Old Mills wheat brad | 221 | Chocolate milk | 11* |
| Wonder wheat bread | 222 | Plain milk | 12* |
| Present chocolate cookies | 311 | White bread | 21* |
| 77 Chocolate cookies | 312 | Wheat bread | 22* |
| Present lemon cookies | 321 | Chocolate cookies | 31* |
| 77 Lemon cookies | 322 | Lemon cookies | 32* |
| Limton black tea beverage | 411 | Black tea beverage | 41* |
| Nestle black tea beverage | 412 | Green tea beverage | 42* |
| Linton green tea beverage | 421 | | |
| Nestle green tea beverage | 422 | | |

**Step-1:**
We are using encoded transaction table1. For example, the item " foremore plain milk " in figure1 is encoded as '122', in which the first digit '1' represents the code 'milk' at level 1, the second digit '2' represents. The flavor 'plain (milk)' at level 2, and the third digit '2' represents the brand 'foremore' at level 3.

Step-2
Suppose that $\gamma$ arbitrarily equals to 5; that means qualified transaction is regarded as a transaction with no more than 5 items purchased in the transaction. Result of this step is a set of qualified transaction as seen in Table 2. Where M={T1,T2,T3, T5,T6 ,T7 ,T8, ,T10,T11,T12}.

Table 2

| Trans_ID | List of items |
|---|---|
| T1 | 222, 411, 211 |
| T2 | 411, 412, 422, 211, 221 |
| T3 | 222, 212, 322, 412 |

| Trans_ID | List of items |
|---|---|
| T5 | 322, 212, 422, 311 |
| T6 | 322, 222, 312, 411 |
| T7 | 311, 412, 221, 311, 322 |
| T8 | 212, 411, 321, 221 |
| T10 | 322, 122 |
| T11 | 222, 111 |
| T12 | 1 11 |

The Qualified set of Transaction (M)

Step-3:
Let k ∈ {1,2,3}, where k is used to store the level number being processed.

Step-4:
The process is started by looking for support of 1-itemsets (p =1) for level k = 1.

Step-5:
Since p ∈ {1,2,3}. It is arbitrarily given $\alpha_p^1 = 2$, $\alpha_p^2 = 1$, $\alpha_p^3 = 0.33$. That means the system just considers support of P-itemsets that is greater than or equal to 1.1, for K=1, and greater than or equal to 1.67 for K=2, and greater than or equal to 1.25, for K=3.

Step-6:
All the items in the transactions are first grouped at level one and their corresponding occurrence are added. Take the items in transaction T1 as an example. The items (222) and (211) are grouped into (2**, 2), representation of grouped data items with their occurrence count is given in table3.

Table-3

| Trans_ID | List of items |
|---|---|
| T1 | (2**, 2) (4**, 1) |
| T2 | (2**, 2) (4**, 3) |
| T3 | (2**, 2) (3**, 1) (4**, 1) |
| T5 | (2**, 1) (3**, 2) (4**, 1) |
| T6 | (2**, 1) (3**, 2) (4**, 1) |
| T7 | (2**, 1) (3**, 3) (4**, 1) |
| T8 | (2**, 2) (3**, 1) (4**, 1) |
| T10 | (1**, 1) (3**, 1) |
| T11 | (1**, 1) (2**, 1) |
| T12 | (1**, 1) |

The Grouped items in transaction

Step-7
Every p-itemset $I^p$ is represented as a fuzzy set on set of qualified transactions M as given by the following results:

**Level k = 1**
**Min_support ($\alpha_p^1$) = 2.0**
**1-Itemset**
**Support of 1-Itemsets**
{1**} = {0.5/T10, 0.5/T11, 1/T12}
{1**} = 2.0
{2**}= {0.67/T1, 0.4/T2, 0.5/T3, 0.25/T5, 0.25/T6, 0.2/T7, 0.5/T8, 0.5/T11}
{2**} = 3.27
{3**}= {0.25/T3, 0.5/T5, 0.5/T6, 0.6/T7, 0.25/T8, 0.5/T10}
{3**} = 2.6
{4**}= {0.33/T1, 0.6/T2, 0.25/T3, 0.25/T5, 0.25/T6, 0.2/T7, 0.25/T8}
{4**} = 2.13

Table- 4

$$L_1^1$$

| 1-itemset | Min_Support |
|---|---|
| {1**} | 2 |
| {2**} | 3.27 |
| {3**} | 2.6 |
| {4**} | 2.13 |

All 1-items considered for further process because their *support* >= $\alpha_1^1$.

**2-Itemset**      **Support of 2-Itemsets**
{1**, 2**} = {0.5/T11 ∧ 0.5/T11} = {0.5/T11}   {1**, 2**} =0.5
{1**, 3**} = {0.5/T10 ∧ 0.5/T10} = {0.5/T10}   {1**, 3**} =0.5
{1**, 4**} = { }   {1**, 4**} = 0
{2**, 3**} = {0.5/T3 ∧ 0.25/T3, 0.25/T5 ∧ 0.5/T5,   {2**, 3**} = 1.2
0.25/T6 ∧ 0.5/T6, 0.2/T7 ∧ 0.6/T7, 0.25/T8 ∧   {2**, 4**} = 1.93
0.5/T8} = {0.25/T3, 0.25/T5, 0.25/T6, 0.2/T7,   {3**, 4**} = 1.2
0.25/T8}

{2**, 4**} = {0.67/T1 ∧ 0.33/T1, 0.4/T2 ∧ 0.6/T2, 0.5/T3 ∧ 0.25/T3, 0.25/T5 ∧ 0.25/T5, 0.25/T6 ∧ 0.25/T6, 0.25/T7 ∧ 0.2/T7, 0.5/T8 ∧ 0.25/T8}
= {0.33/T1, 0.4/T2, 0.25/T3, 0.25/T5, 0.25/T6, 0.2/T7, 0.25/T8}

{3**, 4**} = {0.25/T3 ∧ 0.25/T3, 0.5/T5 ∧ 0.25/T5, 0.5/T6 ∧ 0.25/T6, 0.6/T7 ∧ 0.2/T7, 0.2/T8 ∧ 0.25/T8}
= {0.25/T3, 0.25/T5, 0.25/T6, 0.2/T7, 0.25/T8}

Table- 5

$$L_2^1$$

| 2-itemset | Support |
|---|---|
| {2**, 3**} | 1.2 |
| {2**, 4**} | 1.93 |
| {3**, 4**} | 1.2 |

The {1**,2**}, {1**,3**}, {1**,4**} cannot be considered for further process because their *support* <= $\alpha_2^1$.

**3-Itemset**      **support of 3-Itemsets**
{2**, 3**, 4**} = {0.5/T3 ∧ 0.25/T3 ∧ 0.25/T3,   {2**, 3**, 4**} =1.2
0.25/T5 ∧ 0.5/T5 ∧ 0.25/T5,
0.25/T6 ∧ 0.5/T6 ∧ 0.25/T6,
0.2/T7 ∧ 0.6/T7 ∧ 0.2/T7, 0.5/T8
∧ 0.25/T8 ∧ 0.25/T8}
= {0.25/T3, 0.25/T5, 0.25/T6, 0.2/T7, 0.25/T8}

Table-6

$$L_3^1$$

| 3-itemset | Support |
|---|---|
| {2**, 3**, 4**} | 1.2 |

**Level k =2**
**Min_support ($\alpha_p^2$) = 1.0**

1-Itemset

{21*} = {0.33/T1, 0.2/T2, 0.25/T3, 0.25/T5, 0.25/T8}
{22*} = {0.33/T1, 0.2/T2, 0.25/T3, 0.25/T6, 0.2/T7, 0.25/T8, 0.5/T11}
{31*} = {0.25/T5, 0.25/T6, 0.4/T7}
{32*} = {0.25/T3, 0.25/T5, 0.25/T6, 0.2/T7, 0.25/T8, 0.5/T10}
{41*} = {0.33/T1, 0.4/T2, 0.25/T3, 0.25/T6, 0.2/T7, 0.25/T8}
{42*} = {0.2/T2, 0.25/T5}

Support of 1-itemset

{21*} = 1.28
{22*} = 1.98
{31*} = 0.9
{32*} = 1.7
{41*} = 1.68
{42*} = 0.45

Table-7

$$L_1^2$$

| 1-itemset | min-support |
|---|---|
| {21*} | 1.28 |
| {22*} | 1.98 |
| {32*} | 1.7 |
| {41*} | 1.68 |

The {31*}, {42*} cannot be considered for further process because their *support* <= $\alpha_1^2$.

**2-Itemset**

{21*, 22*} = {0.33/T1 ∧ 0.33/T1, 0.2/T2 ∧ 0.2/T2, 0.25/T3 ∧ 0.25/T3, 0.25/T8 ∧ 0.25/T8}
= {0.33/T1, 0.2/T2, 0.25/T3, 0.25/T8}
{21*, 32*} = {0.25/T3 ∧ 0.25/T3, 0.25/T5 ∧ 0.25/T5, 0.25/T8 ∧ 0.25/T8}
= {0.25/T3, 0.25/T5, 0.25/T8}
{21*, 41*} = {0.33/T1 ∧ 0.33/T1, 0.2/T2 ∧ 0.4/T2, 0.25/T3 ∧ 0.25/T3, 0.25/T8 ∧ 0.25/T8}
= {0.33/T1, 0.2/T2, 0.25/T3, 0.25/T8}
{22*, 32*} = {0.25/T3 ∧ 0.25/T3, 0.25/T6 ∧ 0.25/T6, 0.2/T7 ∧ 0.2/T7, 0.25/T8 ∧ 0.25/T8}
= {0.25/T3, 0.25/T6, 0.2/T7, 0.25/T8}
{22*, 41*} = {0.33/T1 ∧ 0.33/T1, 0.2/T2 ∧ 0.4/T2, 0.25/T3 ∧ 0.25/T3, 0.25/T6 ∧ 0.25/T6, 0.2/T7 ∧ 0.2/T7, 0.25/T8 ∧ 0.25/T8}
= {0.33/T1, 0.2/T2, 0.25/T3, 0.25/T6, 0.2/T7, 0.25/T8}
{32*, 41*} = {0.25/T3 ∧ 0.25/T3, 0.25/T6 ∧ 0.25/T6, 0.2/T7 ∧ 0.2/T7, 0.25/T8 ∧ 0.25/T8}
= {0.25/T3, 0.25/T6, 0.2/T7, 0.25/T8}

Support of 2-Itemsets

{21*, 22*} = 1.03
{21*, 32*} = 0.75
{21*, 41*} = 1.03
{22*, 32*} = 0.95
{22*, 41*} = 1.48
{32*, 41*} = 0.95

Table- 8

$$L_2^2$$

| 2-itemset | min-support |
|---|---|
| {21*, 22} | 1.03 |
| {21*, 41} | 1.03 |
| {22*, 41} | 1.48 |

The {21*, 32*}, {22*, 32*} {32*, 41*} cannot be considered for further process because their *support* <=. $\alpha_2^2$

**3- Itemset**

{21*, 22*, 41*} = {0.33/T1 ∧ 0.33/T1 ∧ 0.33/T1, 0.2/T2 ∧ 0.2/T2 ∧ 0.4/T2, 0.25/T3 ∧ 0.25/T3 ∧ 0.25/T3, 0.25/T8 ∧ 0.25/T8 ∧ 0.25/T8}
= {0.33/T1, 0.2/T2, 0.25/T3, 0.25/T8}
= 1.03

Table- 9

$$L_3^2$$

| 3-itemset | min-support |
|---|---|
| {21*, 22*, 41*} | 1.03 |

**Level k =3**
**Min_support ($\alpha_p^3$) = 0.33**

Support of 1-Itemsets

{211}= {0.33/T1, 0.2/T2}                      {211} = 0.53
{212}= {0.33/T1, 0.25/T5, 0.25/T8}            {212} =0.75
{221} = {0.2/T2, 0.2/T7, 0.25/T8}             {221} =0.65
{222} = {0.33/T1, 0.25/T3, 0.25/T6, 0.5/T11}  {222} =1.33
{411} = {0.33/T1, 0.2/T2, 0.25/T6, 0.25/T8}   {411} =1.03
{412} = {0.2/T2, 0.25/T3, 0.2/T7}             {412} =0.65

Table- 10

$$L_1^3$$

| 1-itemset | min-support |
|---|---|
| {211} | 0.53 |
| {212} | 0.75 |
| {221} | 0.65 |
| {222} | 1.33 |
| {411} | 1.03 |
| {412} | 0.65 |

All itemsets be considered for further process because their

$support >= \alpha_1^3$.

| 2-Itemset | Support of 2-Itemsets |
|---|---|
| {211, 212} = {0.33/T1 ∧ 0.33/T1} = {0.33/T1} | {211, 212} = 0.33 |
| {211, 221} = {0.2/T2 ∧ 0.2/T2} = {0.2/T2} | {211, 221} = 0.2 |
| {211, 222} = {0.33/T1 ∧ 0.33/T1} = {0.33/T1} | {211, 222} = 0.33 |
|  | {211, 411} = 0.53 |
| = 0.25 | {211, 412} = 0.2 |
| {211, 411} = {0.33/T1 ∧ 0.33/T1, 0.2/T2 ∧ 0.2/T2} = {0.33/T1, 0.2/T2} | {212, 221} |
| {211, 412} = {0.2/T2 ∧ 0.2/T2} = {0.2/T2} | {212, 222} = 0.33 |
|  | {212, 411} = 0.58 |
| {212, 221} = {0.25/T8 ∧ 0.25/T8} = {0.25/T8} | {212, 412} = 0 |
| {212, 222} = {0.33/T1 ∧ 0.33/T1} = {0.33/T1} | {221, 222} = 0 |
| {212, 411} = {0.33/T1 ∧ 0.33/T1, 0.25/T8 ∧ 0.25/T8} = {0.33/T1, 0.25/T8} | {221, 411} = 0.25 |
|  | {221, 412} = 0.45 |
| {212, 412} = { } | {222, 411} = 0.58 |
| {221, 222} = { } | {222, 412} = 0.25 |
| {221, 411} = {0.25/T8 ∧ 0.25/T8} = {0.25/T8} | |
| {221, 412} = {0.2/T2 ∧ 0.2/T2, 0.25/T8 ∧ 0.25/T8} = {0.2/T2, 0.25/T8} | |
| {222, 411} = {0.33/T1 ∧ 0.33/T1, 0.25/T6 ∧ 0.25/T6} = {0.33/T1, 0.25/T6} | |
| {222, 412} = {0.25/T3 ∧ 0.25/T3} = {0.25/T3} | |

Table- 11

$$L_2^3$$

| 2-itemset | min-support |
|---|---|
| {211, 212} | 0.33 |
| {211, 222} | 0.33 |
| {211, 411} | 0.53 |
| {212, 222} | 0.33 |
| {212, 411} | 0.58 |
| {221, 412} | 0.45 |
| {222, 411} | 0.58 |

The {211, 221}, {211, 412}, {212, 221}, {212, 412}, {221, 222}, {221, 411}, {222, 412}, {411, 412} cannot be considered for further process because their $support <= \alpha_2^3$.

**3- Itemset**
{211, 212, 222}= {0.33/T1 ∧ 0.33/T1 ∧ 0.33/T1} = 0.33/T1
{211, 212, 411}= {0.33/T1 ∧ 0.33/T1 ∧ 0.33/T1} = 0.33/T1
{211, 212, 412}= { }
{212, 222, 411}= {0.33/T1 ∧ 0.33/T1 ∧ 0.33/T1} = 0.33/T1
{212, 412, 411}= { }
{222, 411, 412}= { }

Table- 12

$$L_3^3$$

| 3-itemset | min-support |
|---|---|
| {211, 212, 222} | 0.33 |
| {211, 212, 411} | 0.33 |
| {212, 222, 411} | 0.33 |

Step-8
Support of each p-itemset is calculated as given in the following results:
Step= 9
From the results as performed by Step-7 and 8, the sets of frequent 1-itemsets, 2-itemsets and 3-itemsets at level 1 are given in Table 4, 5, 6, 7, 8, 9, 10 and 11, 12, respectively.
Step-10
This step is just for increment the value of k in which if $P > \gamma$, then the process is going to Step-12.
Step-11
This step is looking for possible/candidate p-itemsets from $L_{p-1}$ at level k as there is no more candidate itemsets
(as p=3) then go to Step-12. Otherwise, the process is going to Step-5.
Step-12
The step is to calculate every confidence of each possible association rules as follows:

At level k =1

$$Conf(1^{**} \Rightarrow 2^{**}) = \frac{Support(1^{**},2^{**})}{Support(1^{**})} = \frac{0.5}{2} = 0.25$$

.
.
.

Similarly the confidence calculated for itemset

$$Conf(2^{**} \wedge 3^{**} \Rightarrow 4^{**}) = \frac{Support(2^{**},3^{**},4^{**})}{Support(2^{**},3^{**})} = \frac{1.2}{1.2} = 1$$

*Similarly the Confidence calculated for levels k =2&3 are following:*

$$Conf(21^* \wedge 22^* \Rightarrow 41^*) = \frac{Support(21^*,22^*,41^*)}{Support(21^*,22^*)} = \frac{1.03}{1.03} = 1$$

$$Conf(211 \wedge 212 \Rightarrow 222) = \frac{Support(211,212,222)}{Support(211,212)} = \frac{0.33}{0.33} = 1$$

$$Conf(211 \wedge 212 \Rightarrow 411) = \frac{Support(211,212,411)}{Support(211,212)} = \frac{0.33}{0.33} = 1$$

$$Conf(212 \wedge 222 \Rightarrow 411) = \frac{Support(212,222,411)}{Support(212,222)} = \frac{0.33}{0.33} = 1$$

$$Conf(211 \wedge 212 \Rightarrow 222) = \frac{Support(211,212,222)}{Support(211,212)} = \frac{0.33}{0.33} = 1$$

## CONCLUSION

This paper, we have employed fuzzy set concepts, multiple-level taxonomy and different minimum supports to find fuzzy association rules in a given transaction data set. The models works well with problems involving uncertainty in data relationships, which are represented by fuzzy set concepts. The proposed fuzzy mining algorithm can thus generate large itemsets level by level and then derive fuzzy association rules from transaction dataset. The results shown in the example implies that the proposed algorithm can derive the multiple-level association rules under different supports in a simple and effective way.